\documentclass[twocolumn,preprintnumbers,amsmath,amssymb]{revtex4}
\usepackage{braket}
\usepackage{here}
\usepackage{graphicx} 
\usepackage{latexsym,amssymb,amsthm,amsmath,epsfig}
\usepackage{color}

\begin{document}

\title{Scaling Law for Three-Body Collisions in Identical Fermions with $p$-Wave Interactions}
\author{Jun Yoshida$^{1,2}$}
\email{j\_yoshida@ils.uec.ac.jp}
\author{Taketo Saito$^{1,2}$}
\author{Muhammad Waseem$^{1,2}$}
\author{Keita Hattori$^{1,2}$}
\author{Takashi Mukaiyama$^{3}$}
\affiliation{%
$^{1}$\mbox{Department of Engineering Science, University of Electro-Communications, Tokyo 182-8585, Japan}\\
$^{2}$\mbox{Institute for Laser Science, University of Electro-Communications, Chofugaoka, Chofu, Tokyo 182-8585, Japan}\\
$^{3}$\mbox{Graduate School of Engineering Science, Osaka University, Machikaneyama, Toyonaka, Osaka 560-8531, Japan}\\
}
\date{\today}

\begin{abstract}
We experimentally confirmed the threshold behavior and scattering length scaling law of the three-body loss coefficients in an ultracold spin-polarized gas of $^6$Li atoms near a $p$-wave Feshbach resonance. We measured the three-body loss coefficients as functions of temperature and scattering volume, and found that the threshold law and the scattering length scaling law hold in limited temperature and magnetic field regions. We also found that the breakdown of the scaling laws is due to the emergence of the effective-range term. This work is an important first step toward full understanding of the loss of identical fermions with $p$-wave interactions.
\end{abstract}

\maketitle

Three-body collisions, which sometimes appear to present the largest obstacle in achieving exotic many-body quantum states in ultracold atoms~\cite{Iskin1, Iskin2, rich_phase}, provide a key signature of three-body correlation. The ability to tune interatomic interactions using Feshbach resonances has contributed significantly toward clarifying the relation between the three-body collisional properties and scattering length of a system of ultracold atoms with $s$-wave interactions. It has been theoretically proposed and experimentally confirmed that in identical bosons with large $s$-wave scattering lengths, the three-body loss coefficient $L_3$ depends on the scattering length $a_s$ as $L_3 \propto {a_s} ^4$ in a large scattering length regime~\cite{a4_1, a4_2, a4_3, a4_4, a4_5, a4_experiment, a4_experiment2, a4_experiment3,Shotan}. This scaling behavior of $L_3$ has provided the baseline characteristic of three-body losses to mark out the three-body Efimov resonances~\cite{efimov_experiment} and four-body resonances~\cite{Ferlaino} in a system of identical bosons. 

As the three-body physics has been clarified in $s$-wave interacting systems, attention has now turned to the three-body collisional properties arising from enhanced $p$-wave interactions ~\cite{Esry,Suno,Jona,Levinsen}. The $p$-wave interactions can be described by the low-energy expansion of the $p$-wave scattering amplitude $f_{p}(k)$ with an effective range expansion \cite{scattering theory}:
\begin{equation}
f_{p}(k)=\frac{1}{k{\rm cot}\delta_{p}-ik}
,\   k{\rm cot}\delta_{p}=\frac{1}{V_{B} k^2}+k_{\rm e}.
\label{eq:scattering amplitude}
\end{equation}
Here, $\delta_{p}$, $V_B$, $k_{\rm e}$, and $k$ are the $p$-wave scattering phase shift, scattering volume, second coefficient of the effective range expansion, and wave number, respectively.
$V_B$ has the form of a resonance, $V_B = V_{\rm bg}\left(1+\frac{\Delta B}{B-B_{\rm res}}\right) \simeq \frac{V_{\rm bg}\Delta B}{B-B_{\rm res}}$. Here, $V_{\rm bg}$, ${\Delta B}$, and $B_{\rm res}$ are the background scattering volume, resonance width, and resonance magnetic field, respectively, which were experimentally determined previously~\cite{Nakasuji}. In $p$-wave interacting systems, the $k$ dependence of the scattering amplitude becomes more significant than that in $s$-wave interacting systems due to the larger contribution of the effective range term.

At the low-$k$ limit, three-body collisions of indistinguishable fermions with $p$-wave interactions obey the threshold law of $L_3 \propto E^2$, where $E$ is the collision energy~\cite{Esry}. In addition, dimensional analysis assuming the threshold law, indicates that $L_3 \propto {V_B}^{8/3}$ at relatively small $V_B$~\cite{Suno}. This scaling law is true only when the contribution from the effective range is negligible; however, the parameter range in which the scaling behavior actually applies is unclear. Since the first observation of the $p$-wave Feshbach resonance, the atomic losses have been studied extensively~\cite{ENS_loss, JILA_loss, MIT_loss}, the scattering parameters have been determined~\cite{Ticknor, Nakasuji}, $p$-wave molecules have been created ~\cite{JILA_mol, Fuchs, inada, Maier}, and the $p$-wave contacts have been determined~\cite{Luciuk}. Although the three-body collision coefficients have been systematically studied, the scaling behavior has eluded experimental confirmation to date. This is presumably because, in order to satisfy the requirement that the effective range is negligible, the scaling behavior appears only in far-detuned regimes from the $p$-wave Feshbach resonance, such as the limited scattering volume regime~\cite{Suno}, where the atomic lifetime due to the three-body loss is similar to a one-body atomic lifetime. 

In this study, we measured the three-body loss coefficients as functions of temperature and magnetic field in an ultracold spin-polarized gas of $^6$Li atoms near a $p$-wave Feshbach resonance. We experimentally confirmed the threshold behavior of $L_3 \propto E^2$ and the scattering length scaling of the three-body loss coefficient, $L_3 \propto |V_B|^{8/3}$. We focused on the large-detuning magnetic field regime, where the scaling law is expected to hold. We also worked at the low-temperature regime, where the threshold behavior still holds; however, the $^6$Li atoms were maintained at a temperature above the Fermi temperature, and therefore, a Gaussian spatial density profile for the atomic cloud could be assumed. One advantage of using $^6$Li as opposed to $^{40}$K is that the atomic loss in $^6$Li is caused only by three-body losses, unlike the case for $^{40}$K, because the lowest spin state of $^6$Li is not a stretched state; therefore, we can work with the Feshbach resonances in the lowest-energy state, where dipolar losses do not occur~\cite{JILA_mol,Kurlov}.


A gas of $^6$Li atoms was prepared by an all-optical means as described in \cite{Nakasuji}. We trapped two-component fermionic atoms of $^6$Li in the $\ket{1} \equiv \ket{F=1/2, m_F=1/2}$ and $\ket{2} \equiv \ket{F=1/2, m_F=-1/2}$ states, and performed evaporative cooling at a magnetic field of 300~G. The gas was cooled to approximately $T/T_{\rm F}\simeq 1$, where $T$ is the temperature of the $^6$Li atoms, and $T_{\rm F}$ is the Fermi temperature. We limited our experiment to gases with temperatures of $T/T_{\rm F}\simeq 1$ to ensure that the atomic density profile could be described well by a Gaussian function. After the evaporative cooling, we irradiated the atoms in the $\ket{2}$ state with resonant light to remove them from the trap, in order to prepare a spin-polarized Fermi gas of atoms in the $\ket{1}$ state. The magnetic field was then scanned quickly below the $p$-wave Feshbach resonance for atoms in the $\ket{1}$ state, which are located at a magnetic field of 159~G. Next, the magnetic field was altered to the value at which the loss of the atoms was measured. We calculated the number of atoms remaining after various holding times to obtain the decay curve of the atoms in particular magnetic fields. In our experiment, we limited the loss of atoms to less than $30 \%$ of the initial number of atoms to suppress the effect of deformation of the momentum distribution.

\begin{figure}[b]
\includegraphics[width=2.6 in]{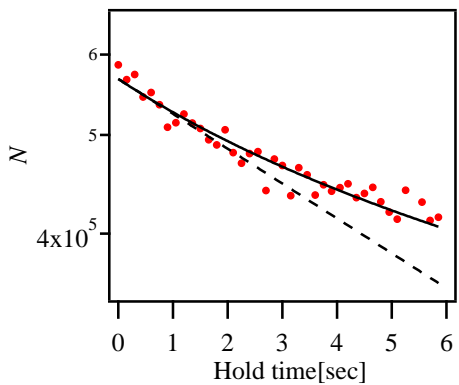}
\caption {Typical atomic decay plot. Each data point is the average of three measurements. The solid curve shows the result of fitting with Eq.~\ref{eq:rate}, and the dashed line indicates exponential decay.} 
\label{example_plot}
\end{figure} 

The decay of the number of atoms is described by the rate equation for the atomic density as 
\begin{eqnarray}
\dot{N}/N &=& -L_{3} (T, B) \langle{ n^{2}\rangle},
\label{eq:rate}
\end{eqnarray}

\noindent where $\langle{ n^{2}\rangle}$, $N$, and $L_3 (T, B)$ is mean square density, total atom number, and trap-averaged three-body loss coefficient as a function of the temperature $T$ and magnetic field $B$, respectively. $L_{3} (T, B)$ can be described by averaging $L_3 (E, B)$ over the collision energy $E$ assuming a Maxwell--Boltzmann distribution as \cite{Suno, Zhenhua}
\begin{equation}
L_3 (T, B) =\frac{1}{2(k_{\rm B}T)^3} \int L_3 (E, B) E^2 e^{-E/k_{\rm B}T} {\rm d}E.
\end{equation}
$L_{3} (T, B)$ was measured by fitting the decay curve to Eq. \ref{eq:rate}. The accuracy of $L_{3} (T, B)$ is fairly affected by the accuracy of the atom number counting in the absorption imaging. However, we have confirmed that the systematic uncertainty of the atom number counting does not affect the observation of scaling behavior \cite{N_error}.
Our experimental setup had an atomic trap lifetime of 100 s, which was determined by background gas collisions. Therefore, we limited our measurements to the three-body loss rate, which is at least twice as fast as the one-body loss rate.
The decay curve was measured three times for each experimental condition; one decay curve comprised 40 shots with various holding times. A typical atomic decay plot is shown in Fig.~\ref {example_plot}. 
The solid curve indicates the result of fitting with Eq.~\ref{eq:rate} to obtain $L_{3}$. The dashed line indicates the exponential decay of $N=N_{0} e^{-\Gamma t}$ with $\Gamma=L_{3} \times \langle{ n^{2}\rangle}_{t=0} $. It is clearly seen that the atomic decay curve follows the non-exponential decay behavior set by Eq.~\ref{eq:rate}, due to the three-body loss processes, which are proportional to the mean square density.

\begin{figure}[b]
\includegraphics[width=3.4 in]{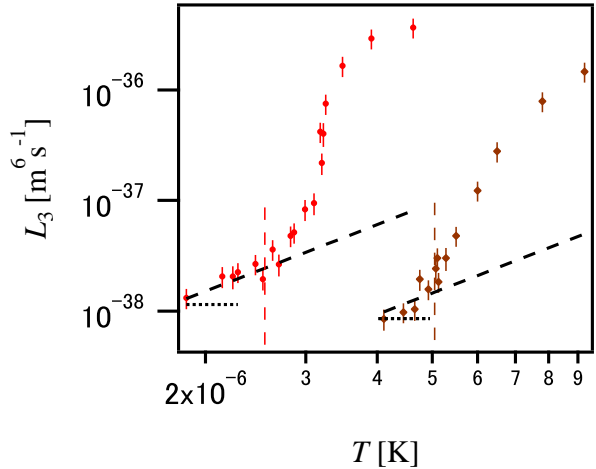}
\caption {Three-body loss coefficients vs. temperature at two different magnetic field detunings. The red circles and brown rhombi indicate the data at magnetic field detunings of 0.22 and 0.45~G, respectively. The error bars indicate the statistical error arising from the statistical errors in the number of atoms ($15\ \%$) and the temperature ($5\ \%$). 
The black dotted horizontal lines indicate the point where the three-body loss rate is twice as fast as the one-body loss rate.
The black dashed vertical lines show the quadratic dependence of $L_3$ on the temperature, which is consistent with the threshold behavior predicted by Esry et al.~\cite{Esry}. The red and brown dashed vertical lines indicate the temperature at which $k_{\rm e} {k_T}^2 V_B = 0.095$ for the data shown in each color. } 
\label{Tempdep}
\end{figure} 

{Fig.~\ref{Tempdep} shows a double logarithmic plot of the three-body loss coefficients $L_3$ as a function of the atomic temperature at magnetic fields detuned by 0.22 G (red circles) and 0.45~G (brown rhombi) toward the Bardeen--Cooper-Schrieffer (BCS) side of the resonance, which correspond to scattering volumes of $(-232~a_0)^3$ and $(-183~a_0)^3$, respectively, where $a_0$ is the Bohr radius. $L_3$ increases gradually in the low-temperature region and rises rapidly above a certain temperature. According to the threshold behavior predicted by Esry et al.~\cite{Esry}, the three-body loss coefficient has a quadratic dependence on the collision energy. Because our measurements are limited to the temperature region in which the Maxwell--Boltzmann distribution applies, the thermal averaged $L_3$ exhibits a threshold behavior of $L_3 \propto T^2$. The black dashed vertical lines in Fig.~\ref{Tempdep} show the fitting results for $L_3 \propto T^2$ with a prefactor as a free parameter in the low-temperature region. $L_3$ shows a quadratic dependence, as expected from the threshold behavior. In the high-temperature region, $L_3$ deviates from the threshold behavior quite rapidly. The red and brown dashed vertical lines indicate the temperature at which $k_{\rm e} {k_T}^2 V_B = 0.095$ for the data shown in each color. Here, $k_T$ is the mean wave number of atoms, which is defined as $k_T = \sqrt{3mk_{\rm B} T/2 \hbar^2}$. The dimensionless quantity $k_{\rm e} {k_T}^2 V_B$ is the ratio of the second term to the first term in the second equation in Eq.~(1) and indicates the contribution of the effective range term to the collisional phase shift. The temperature dependence of $L_3$ shows that the threshold law breaks down at the same value of $k_{\rm e} {k_T}^2 V_B$ at two different magnetic field detunings. This fact indicates that the effective range term plays an important role in the breakdown of the threshold law.

\begin{figure}[b]
\includegraphics[width=3.4 in]{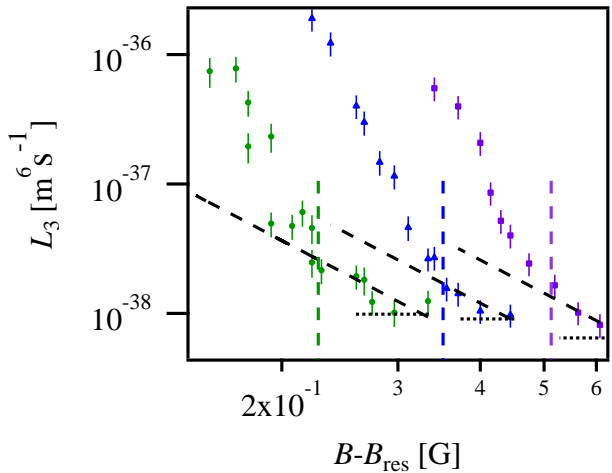}
\caption {Three-body loss coefficients vs. magnetic field detuning for three temperatures. The green circles, blue triangles, and purple squares show the data at 2.7, 3.9, and 5.7~$\mu$K, respectively. The error bars indicate the statistical error arising from the statistical errors in the number of atoms ($15\ \%$) and the temperature ($5\ \%$). 
The black dotted horizontal lines indicate the point where the three-body loss rate is twice as fast as the one-body loss rate.
The black dashed vertical lines show the $V_B^{8/3}$ dependence of $L_3$, as predicted by Suno et al.~\cite{Suno}. The green, blue, and purple dashed vertical lines indicate the magnetic field detuning points at which $k_{\rm e} {k_T}^2 V_B = 0.095$ for the data shown in each color.}
\label{magdep}
\end{figure} 

Next, we discuss the magnetic field dependence of the three-body loss coefficient. Fig.~\ref{magdep} shows $L_3$ vs. the magnetic field detuning $B-B_{\rm res}$ for measurements at temperatures of 2.7~$\mu$K (green circles), 3.9~$\mu$K (blue triangles), and 5.7~$\mu$K (purple squares). Here, the magnetic field is tuned from a detuning of 0.15~G to a detuning of 0.6~G toward the BCS side of the resonance, which corresponds to tuning the scattering volume from $(-263~a_0)^3$ to $(-166~a_0)^3$. From the large detuning to the small detuning, $L_3$ increases gradually, and it rises rapidly at certain magnetic field detunings; this behavior is similar to the temperature dependence of $L_3$ (as shown in Fig.~\ref{Tempdep}). From the dimensional analysis taking into account the threshold behavior, $L_3$ is expected to vary with the $p$-wave scattering volume as $L_3 \propto {V_B}^{8/3}$~\cite{Suno}, assuming that the effective range contribution is negligible. In Fig.~\ref{magdep}, the black dashed lines show the results of fitting the ${V_B}^{8/3}$ dependence in a relatively large detuning region for the three datasets. The experimental results are clearly consistent with the scattering length scaling law of $L_3 \propto {V_B}^{8/3}$ in certain magnetic field detuning regions. $L_3$ deviates rapidly in the region relatively close to the Feshbach resonance. The green, blue, and purple dashed vertical lines indicate the magnetic field detuning points at which $k_{\rm e} {k_T}^2 V_B = 0.095$} for the data shown in each color. Notably, the three sets of data deviate from the scattering length scaling behavior at the points corresponding to $k_{\rm e} {k_T}^2 V_B = 0.095$, which share the same critical value with the temperature dependence. This fact seems to suggest that the effective range term again plays an important role in the breakdown of the scattering length scaling law.

\begin{figure}[b]
\includegraphics[width=3.4 in]{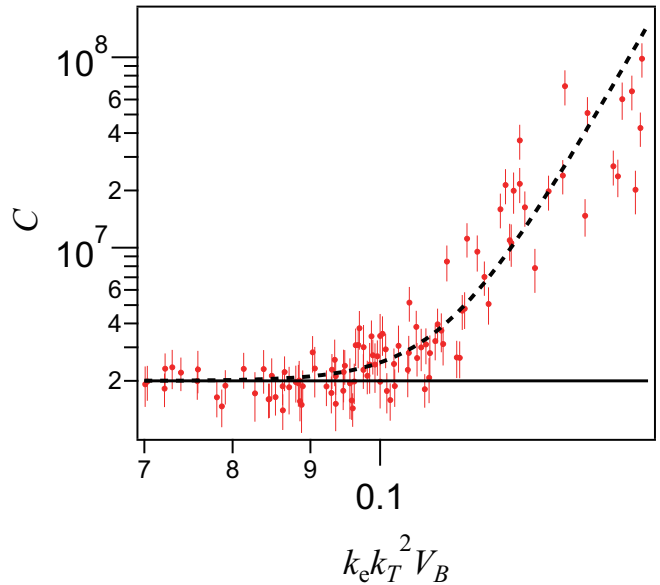}
\caption {Dimensionless quantity $C$ vs. $k_{\rm e} {k_T}^2 V_B$. The black solid line indicates the constant value at which scaling behavior is exhibited. The black dashed curve shows the results of fitting with modified Eq.~\ref{eq:scaling} with $\beta = 9$ and $\gamma = 14$.}
\label{dimensionless}
\end{figure} 

Because the three-body loss coefficient obeys the threshold law and scattering length scaling law, $L_3$ can be described by 
\begin{equation}
L_3 = C \frac{\hbar}{m}\times {k_{T}}^4 V_B^{8/3},
\label{eq:scaling}
\end{equation}

where $C$ is a dimensionless quantity whose magnitude reflects the original strength of the three-body loss of the Feshbach resonance under consideration, which is determined by the coupling between the closed channel and the tightly bound dimer state. $C$ must be constant as long as the threshold law and scattering length scaling law hold. In Fig.~\ref{dimensionless}, the $C$ values for all the data shown in Figs.~\ref{Tempdep} and \ref{magdep} and additional data that fall into the same parameter region are plotted as a function of the dimensionless parameter $k_{\rm e} {k_T}^2 V_B$. The entire data clearly collapse onto a single monotonic curve starting from a constant value in the smaller $k_{\rm e} {k_T}^2 V_B$ region. 
The data show that the whole scaling law holds below $k_{\rm e} {k_T}^2 V_B \sim 0.095$ where $C$ shows constant behavior and deviates at higher $k_{\rm e} {k_T}^2 V_B$ values.

Next, we examine the deviation of the data from the scaling behavior observed at $k_{\rm e} {k_T}^2 V_B > 0.095$. If we assume that the deviation is due to the contribution of the effective range term, $L_3$ can be described in terms of the expansion of the small quantity $k_{\rm e} {k_T}^2 V_B$. Consequently, $C$ can be modified as $C = C_0 \{1 + {(\beta \times k_{\rm e} {k_T}^2 V_B)}^\gamma\}$,
where $\beta$ and $\gamma$ are free parameters, and $C_0 = 2.0 \times 10^6$. The dashed curve in Fig.~\ref{dimensionless} shows the best-fit result of the function above with $\beta=9$ and $\gamma=14$. This indicates that $L_3$ increases quite rapidly as a function of $k_{\rm e} {k_T}^2 V_B$. 

The departure from the scaling law was theoretically studied by Suno et al.~\cite{Suno}. Their calculation clearly shows a steep increase in $L_3$ with increasing $V_B$, similar to that observed in the current experiment. They attribute the steep increase in $L_3$ in the large $V_B$ region to the effect of the resonant tunneling of two free colliding atoms into the bound dimer state. Since $k_{\rm e} {k_{T}}^{2} V_B$ quantifies how close the dimer bound state energy level is to the average collision energy of two atoms, our finding in this paper that the breakdown of the scaling law is determined by $k_{\rm e} {k_{T}}^{2} V_B$ is consistent with the previous discussion by Suno~\cite{Suno}. The full understanding of this near-resonant behavior of three-body loss is an important future challenge that needs to be investigated both experimentally and theoretically.


In conclusion, we experimentally confirmed the threshold behavior and scattering length scaling law of three-body loss coefficients in an ultracold spin-polarized gas of $^6$Li atoms near a $p$-wave Feshbach resonance. We measured the three-body loss coefficients as functions of temperature to confirm the quadratic dependence of $L_3$ on temperature in the low-temperature region. We also measured the scattering volume dependence of $L_3$ to confirm the scattering length scaling law of $L_3 \propto {V_B}^{8/3}$ in the small $V_B$ region. Moreover, we experimentally confirmed that the scaling laws break down at $k_{\rm e} k^2 V_B \sim 0.095$, indicating that the effective range term comes into play in the three-body loss. After the breakdown of the scaling laws, we observed a steep increase in the three-body loss coefficient with $k_{\rm e} k^2 V_B$. Further experimental and theoretical studies are required to obtain a full understanding of $L_3$ in the near-resonant region, where a $p$-wave superfluid is expected to appear.


We thank Dr. Zhenhua Yu for fruitful discussions and Dr. Shinsuke Haze for his support throughout the experiment. This work was supported by a Grant-in-Aid for Scientific Research on Innovative Areas (Grant No. 24105006) and a Grant-in-Aid for Challenging Exploratory Research (Grant No. 17K18752). MW acknowledges the support of a Japanese government scholarship (MEXT).


\begin{thebibliography}{90}
\bibitem{N_error} We hypothetically took the atom number counting efficiency of only 30\%, and analyzed all the data. The values of $L_3$ changed quite a bit, but the threshold behavior and the scaling behavior of $L_3$ were not affected by the imperfection of atom number counting.
\end{thebibliography}
\end{document}